\documentclass[
reprint,
superscriptaddress,
bibnotes,
amsmath,amssymb,
aps,
prb,
floatfix
]{revtex4-2}

\usepackage{graphicx}
\usepackage{dcolumn}
\usepackage{bm}
\usepackage{xcolor}
\usepackage{hyperref}
\hypersetup{
    colorlinks=true,
    linkcolor=blue,
    citecolor=blue,
    urlcolor=blue,
    pdftitle={Electronic sound in multiband metals},
}
\usepackage{braket}
\usepackage[normalem]{ulem}
\usepackage[none]{hyphenat}

\newcounter{para}

\begin{document}

\title{Electronic Sound and Diffusion in Disordered Multiband Metals}

\author{Miguel-Ángel Sánchez-Martínez}
 \affiliation{School of Physics, Tyndall Avenue, Bristol, BS8 1TL, United Kingdom}
\author{Blaise Gout\'eraux}%
\affiliation{CPHT, CNRS, École polytechnique, Institut Polytechnique de Paris, 91120 Palaiseau, France}
\author{Louk Rademaker}%
\affiliation{Department of Quantum Matter Physics, University of Geneva, CH-1211 Geneva, Switzerland}
\affiliation{Lorentz Institute for Theoretical Physics, Leiden University, PO Box 9506, 2300 Leiden, The Netherlands}
\author{Felix Flicker}%
\affiliation{School of Physics, Tyndall Avenue, Bristol, BS8 1TL, United Kingdom}

\date{\today}

\begin{abstract}
Multiband metals can host acoustic plasmons: gapless collective charge excitations involving out-of-phase motion of electrons in different bands, sometimes referred to as `demons'. Using a hydrodynamic description valid at high temperatures, and a microscopic description with vertex corrections valid at low temperatures, we show, upon including disorder inevitably present in real materials, that the mode frequency becomes purely imaginary, resulting in diffusive behaviour over a finite range of small wave vectors $q\le q_c$. This framework provides a natural, parameter-free explanation for the anomalous nonlinear dispersion recently reported for the demon in Sr$_2$RuO$_4$, which deviates from the gapless behaviour predicted by the random phase approximation. Our results establish demons as fundamentally distinct from previously known acoustic excitations, such as phonons and photons, which exhibit inertial dynamics protected by symmetries or gauge redundancies.
\end{abstract}

\maketitle

\section{Introduction}

Hydrodynamics is a universal effective theory capturing the long wavelength, late time relaxational dynamics of globally conserved quantities in interacting thermal systems \cite{landau1987fluid,chaikinlubensky1995}. While initially devised to describe classical fluids, it has been successfully employed to model quantum fluids such as the quark-gluon plasma produced in heavy-ion collisions \cite{Berges:2020fwq}, ultracold atomic gases \cite{Brown:2019vef}, or  the 2D electron fluid in ultraclean graphene \cite{lucas_hydrodynamics_2018,fritz2023hydrodynamicelectronictransport}. While the hydrodynamics of electron fluids was long anticipated theoretically \cite{gurzhi}, its experimental relevance for metals was hindered by the inevitable presence of disorder in real materials, which instead enhances ballistic electronic transport \cite{lucas_hydrodynamics_2018,fritz2023hydrodynamicelectronictransport}. On the other hand, strong electronic correlations lead to fast local thermalization (with holographic quantum matter as a paradigmatic example \cite{hartnoll_holographic_2018,Liu:2020rrn}), and an enhanced hydrodynamic regime \cite{Zaanen:2018edk}, as was realized in graphene near the charge neutrality point \cite{Bandurin_2016,Crossno_2016}.

These ideas have become timely for correlated metals at finite electronic density following the recent report of an acoustic electronic collective mode in Sr$_2$RuO$_4$~\cite{husain_pines_2023} dubbed Pines' demon. This is a 3D acoustic plasmon arising as a collective mode of electrons between different bands~\cite{pines_demon_original,nozieres_electron_1958,gutfreund_acoustic_1973,ruvalds_are_1981}. The conditions for its existence have been studied for decades~\cite{pines_demon_original,gutfreund_acoustic_1973,ruvalds_are_1981}, but the experimental evidence of its existence was only obtained in 2024~\cite{husain_pines_2023}. The data show certain inconsistencies with the original predictions~\cite{pines_demon_original,nozieres_electron_1958,gutfreund_acoustic_1973,ruvalds_are_1981}: while the dispersion of an acoustic mode intercepts the origin --- having zero energy at zero momentum --- the measured dispersion of Pines' demon in Sr$_2$RuO$_4$ does not, and the low energy behaviour remains experimentally challenging to resolve. Crucially, real materials are never perfectly clean: at the energy and momentum scales of the momentum-resolved electron energy loss spectroscopy (m-EELS) measurements used to observe Pines' demon~\cite{husain_pines_2023}, even small amounts of disorder can introduce notable momentum relaxation and thereby qualitatively reshape collective dynamics. 

Microscopically, standard theoretical approaches based on the Random Phase Approximation (RPA) do not capture the observed behaviour~\cite{kliewer_lindhard_1969,kugler_theory_1975,mahan_many-particle_2000,vos_rpa_2025,schultz_optical_2025}. Moreover, commonly used ad hoc vertex corrections to treat disorder, such as Hubbard~\cite{mahan_many-particle_2000} or Mermin~\cite{mermin_lindhard_1970} schemes, can become unphysical once momentum relaxation is introduced. A controlled treatment of disorder-induced vertex corrections is therefore required to obtain a consistent dielectric response and to interpret the demon's dispersion. 

\begin{figure}
    \centering
    \includegraphics[width=
    0.7\linewidth]{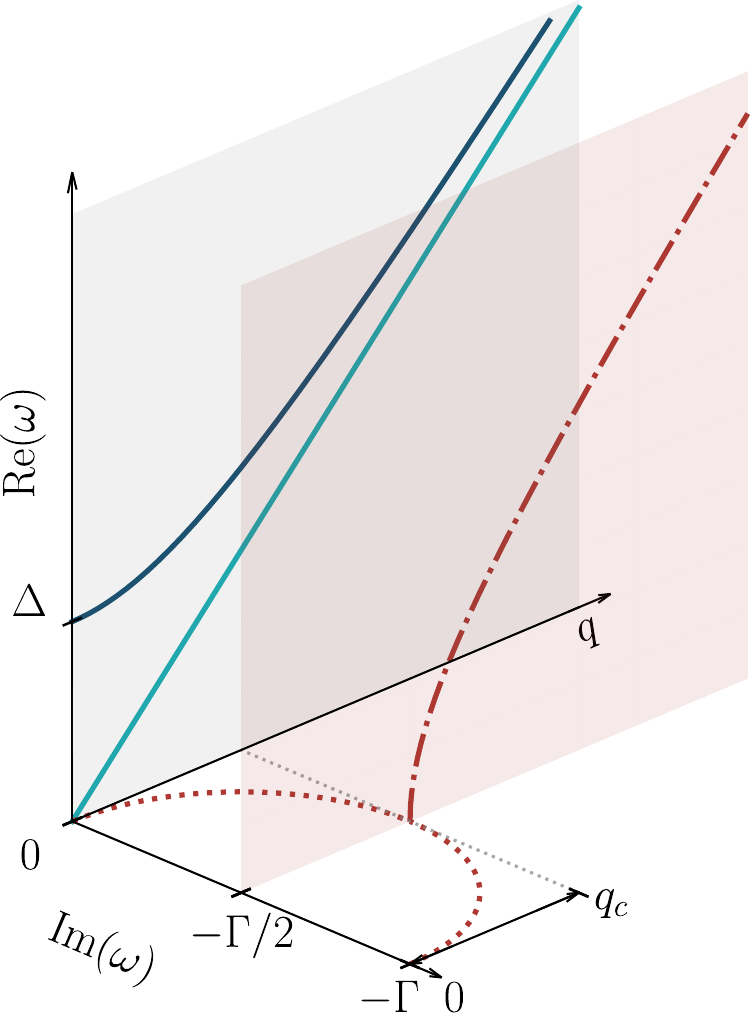}
    \caption{Gapped (dark blue) and gapless (turquoise) dispersions are well known, corresponding to massive and massless quasiparticles or collective modes. Here we report that electrons in multiband metals generically host a different dispersion (red; Eq.~\eqref{eqn:hydro_dispersion}) in the presence of momentum relaxing disorder $\Gamma$. Solid (dotted) lines indicate purely real (imaginary) dispersions; dot-dashed indicate complex dispersions.}
\label{fig:dispersion_relations}
\end{figure}

Here we show that the presence of any amount of momentum relaxation -- always present in real materials -- fundamentally alters its dispersion relation. The result (Fig.~\ref{fig:dispersion_relations}) is well known in studies of diffusive dynamics, but is unusual within the Fermi liquid regime. Indeed, so far as the authors are aware, the 2024 m-EELS study on Sr$_2$RuO$_4$ was, with hindsight, the first observation of such a microscopic dispersion~\cite{husain_pines_2023}.

In this general setting, we derive the dispersion in both momentum-relaxed hydrodynamics and a microscopic quantum theory incorporating disorder vertex corrections. We find perfect analytical agreement between the two, providing a direct link between hydrodynamic modes and the quasiparticle picture, connecting the notion of wave vector in hydrodynamics with that of quasiparticle momentum in the microscopic theory~\footnote{Here the notion of quasiparticle in the microscopic theory is not that of the electrons satisfying the separation of scales between quasiparticle scattering time and transport or total momentum relaxation time in hydrodynamics, but that of Pines' demon, the emergent quasiparticle of anti-phase interband electron oscillations.} across the observed temperature range 30 K to 300 K.

Specialising to Sr$_2$RuO$_4$ we quantitatively predict the dispersion of Pines' demon and compare it with the experimental results in Ref.~\cite{husain_pines_2023}, obtaining a remarkable agreement. Our analysis provides a direct link from the momentum-relaxation rate $\Gamma$---which also controls the resistivity and bad-metallic transport of Sr$_2$RuO$_4$~\cite{stricker_optical_2014,wang_separated_2021}---to the m-EELS response\cite{husain_pines_2023}. More broadly, it implies that acoustic, non-Goldstone collective modes in multiband disordered metals will generically acquire a dispersion as in Fig.~\ref{fig:dispersion_relations} due to momentum relaxation, with direct implications for strange metals and the superconductivity mechanisms in exotic, multiband, and high-Tc materials. 

\section{Results}

The central object by which to study Pines' demon's dispersion is the dielectric function $\epsilon(q,\omega)$ as a function of electronic momentum transfer~\footnote{Unlike in classical hydrodynamic settings, here $q=k-k'$ is the physical momentum transferred to a sample by an external electron scattering from momentum $k$ to $k'$.} $q$ and angular frequency $\omega$~\cite{pines_demon_original,gutfreund_acoustic_1973,ruvalds_are_1981}, which encodes how a material screens charges. Collective charge modes are obtained by solving $\epsilon(q,\omega)=0$, corresponding to poles in the physical response function $\epsilon^{-1}$. For optical plasmons at small $q$, the plasmon is an eigenmode of the Hamiltonian and satisfies ${\rm Im}\,\epsilon(q,\omega)=0$~\cite{mahan_many-particle_2000,bruus_many-body_2004}, so the dispersion can be read off from ${\rm Re}\,\epsilon(q,\omega)=0$. Acoustic plasmons in 3D are qualitatively different. They generally lie inside the particle-hole continuum, meaning they acquire intrinsic damping with ${\rm Im}\,\epsilon(q,\omega)\neq 0$. Their dispersions must be extracted from the complex-frequency solutions of $\epsilon(q,\omega)=0$ with $\omega\in\mathbb{C}$. We proceed to present this calculation using both hydrodynamic and microscopic theories.

\subsection{Hydrodynamics\label{sec:hydrodynamics}}

Pines' demon is an acoustic electronic collective mode that can be viewed as an out-of-phase combination of two electronic densities, producing a neutral `sound'-like excitation. To describe its long-wavelength dynamics in the presence of disorder we use momentum-relaxed hydrodynamics~\cite{chaikinlubensky1995} improved to account for slow momentum relaxation ~\cite{Davison:2014lua,gouteraux_beyond_2024}:
\begin{align}
\label{eqn:hydro_conservation_equations}
    \partial_t \pi^i +n\partial^i\mu&=-\Gamma n v^i-e \frac{n}{q}E^i,\\ \nonumber
    \partial_tn + \partial_i (n v^i)&=0.\nonumber
\end{align}
Here, $\pi^i$ is the total momentum density along the $i^{\textrm{th}}$ spatial direction, \emph{i.e.}~the sum over the momenta of individual long-lived particles when present in the system.  $\pi^i=n v^i$ is the expectation value of the momentum density in the local thermal state, where $v^i$ is the collective velocity of the electronic fluid, and $n$ is the density. $\partial_t$ $(\partial_j)$ is the partial derivative with respect to time (spatial coordinate), $\mu$ is the chemical potential, and $E^i$ the external electric field. $\Gamma$ is the momentum relaxation rate. The top line in  Eq.~\eqref{eqn:hydro_conservation_equations} captures the flux of momentum and the effect of momentum relaxation, while the second line describes the conservation of density. We have assumed the electronic fluid is Galilean invariant. The charge current is obtained by the relation $j^i=-nev^i$. We have neglected viscous corrections as well as temperature and energy fluctuations, which do not play a role here. It is well-known that hydrodynamics can be recovered from kinetic theory for a Fermi-Dirac system: for instance, see \cite{Belitz:2021dlt,lucas_hydrodynamics_2018,fritz2023hydrodynamicelectronictransport}.

Because the acoustic mode described by Eq.~\eqref{eqn:hydro_conservation_equations} is charge neutral, it is not affected by Coulomb interactions. Its dispersion is therefore obtained directly from the zeroes of the susceptibility (see Methods), which obey
\begin{align}
    -\omega^2-i\Gamma\omega+q^2 v_s^2 = 0,
\end{align}
where $v_s=\sqrt{n/(\partial n/\partial\mu)_T}$ is the speed of sound in the fluid. This leads to a dispersion relation of the form
\begin{align}
\label{eqn:hydro_dispersion}
    \omega_{\pm}(q) = -i\frac{\Gamma}{2}\pm\sqrt{q^2 v_s^2 -\frac{\Gamma^2}{4}}.
\end{align}
This dispersion relation corresponds to a neutral mode in a system with finite momentum relaxation $\Gamma$~\cite{Davison:2014lua}. This sets the critical momentum $q_c=\frac{\Gamma}{2v_s}$. This in turn defines three regimes in the dispersion: for $q<q_c$ the dispersion becomes purely imaginary (see Fig.~\ref{fig:dispersion_relations}), and the imaginary part evolves from $\mathrm{Im}(\omega)=0$ ($\mathrm{Im}(\omega)=-\Gamma$) at $q=0$ for the positive (negative) branch of the dispersion to $\mathrm{Im}(\omega)=-\Gamma/2$ for $q=q_c$; for $q \gtrsim q_c$ the dispersion acquires a real part, and the effects of $q$ and $\Gamma$ are comparable, leading to a crossover regime where the behaviour of the real part of the dispersion is not linear; finally, for $q\gg q_c$ the real part of the dispersion tends to a linear behaviour $\omega(q\gg q_c)\sim q$, while the imaginary part is fixed at $\mathrm{Im}(\omega)=-\Gamma/2$ for $q\geq q_c$. 

To obtain the dispersion of Pines' demon in Sr$_2$RuO$_4$, we need estimates of $v_s$ and $\Gamma$ in Eq.~\eqref{eqn:hydro_dispersion}. Using a minimal two-band model fitted to a three-band tight-binding model~\cite{zabolotnyy_renormalized_2013}, we estimate $v_{s}^{30\,\mathrm{K}}=0.1021\,\mathrm{eV}\,\text{\AA}$ and $v_{s}^{300\,\mathrm{K}}=0.1472\,\mathrm{eV}\,\text{\AA}$. We estimate $\Gamma$ using a Drude fit to optical conductivity measurements~\cite{stricker_optical_2014}, as $\Gamma({300\,\mathrm{K}})=51\,\mathrm{meV}$. At $T=30\,\mathrm{K}$, however, a single-$\Gamma$ description is known to be insufficient~\cite{wang_separated_2021}. A two-Drude fit captures well the behaviour at low frequencies~\cite{stricker_optical_2014}, resulting in a momentum relaxation rate $\Gamma(30 {\rm K})= 35.67$~meV. The fitting procedure is described in detail in the Methods section for $T=300$~K and $T=30$~K.

\begin{figure*}[h!]
    \centering
    \includegraphics[width=0.8\linewidth]{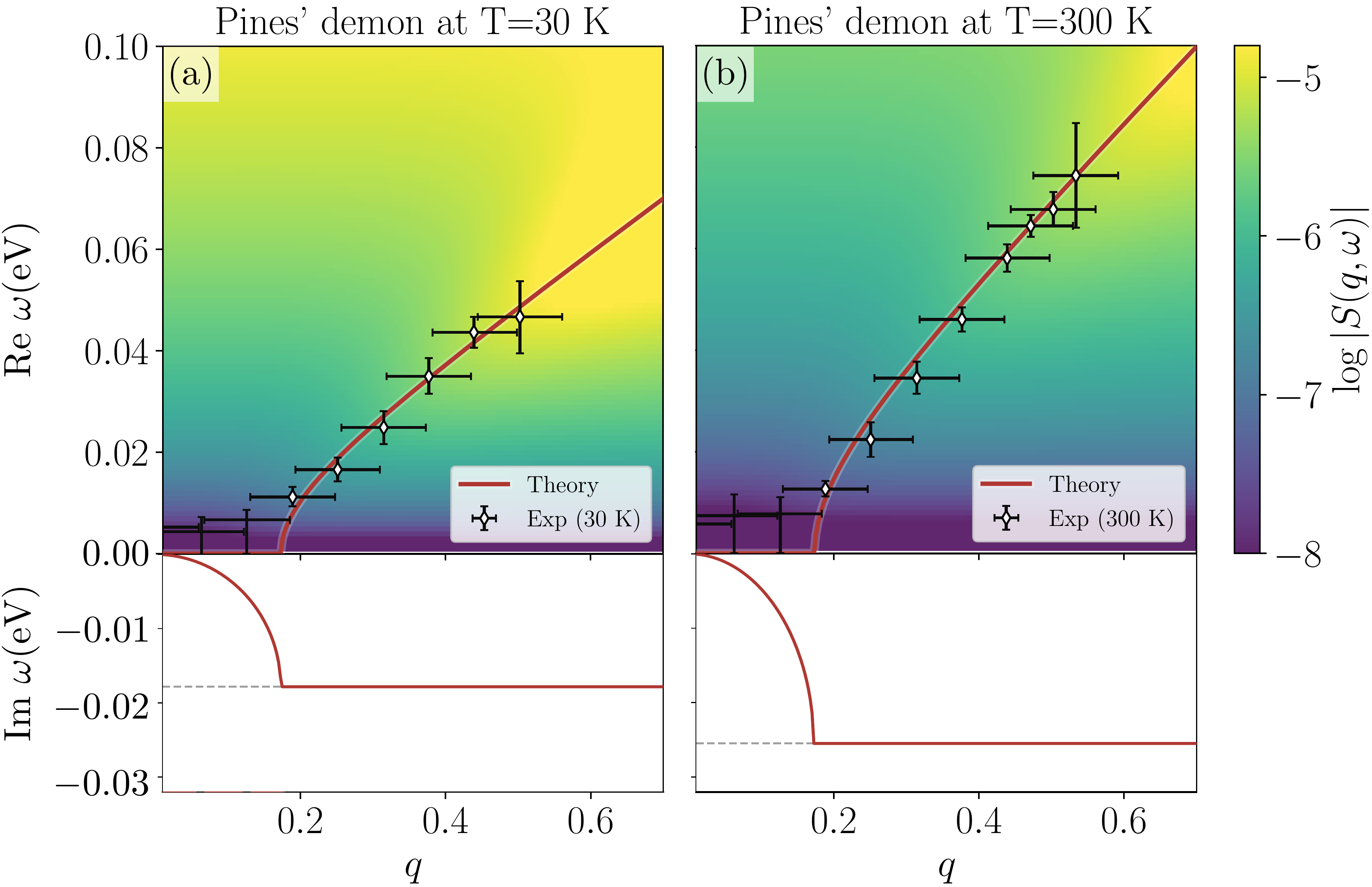}
    \caption{Dispersion of Pines' demon in Sr$_2$RuO$_4$. The theoretical predictions based on the hydrodynamic and microscopic theories for the acoustic plasmon dispersion (solid red) are indistinguishable by eye, and are compared to the experimental m-EELS measurements (black and white diamonds) presented in Ref.~\cite{husain_pines_2023}. Note that the lowest-$q$ data are only error bars, matching our prediction that the demon is purely dissipative at low momentum. We show the loss function $S(q,\omega)$ as a heatmap to aid the visualisation of the acoustic mode as a brighter region in the $(q,\omega)$ plane. We constrain the one free parameter in our model, the momentum relaxation rate $\Gamma$, by fitting it to optical conductivity measurements~\cite{stricker_optical_2014}. The lower panels in both figures present the imaginary component of the dispersion with $q$, showing the evolution of the positive branch in Eq.~\ref{eqn:hydro_dispersion} from $\mathrm{Im}~\omega=0$ for $q=0$ to a finite value $\mathrm{Im}~\omega=-\Gamma/2$ (dashed gray lines) for $q=q_c$ in each case.}
\label{fig:main_results_dispersion}
\end{figure*}

These parameters yield the dispersion shown in Fig.~\ref{fig:main_results_dispersion}. We find remarkably good agreement between the resulting dispersion and the m-EELS measurements reported in Ref.~\cite{husain_pines_2023}. Our theory has only one free parameter, $\Gamma$, which we constrain using optical conductivity data~\cite{stricker_optical_2014} to give an overall parameter-free model. The mode velocity is captured accurately, and the size of $q_c$ is also reproduced without any input from the m-EELS data. It is important to note that the three lowest momentum points in the experiment were null measurements: no demon was seen, and the error bars are estimates based on the experimental resolution~\cite{husain_pines_2023}. They are compatible with zero frequency, matching our prediction. Furthermore, our prediction lies within the error bars of all data points at all measured temperatures, lending weight to the momentum relaxation hypothesis.

\subsection{Microscopic quantum calculation\label{sec:microscopic}}

Accurate incorporation of momentum relaxation into microscopic response functions has been sought for decades owing to its importance to realistic materials modelling. Often this is done phenomenologically by modifying RPA-based dielectric functions~\cite{mermin_lindhard_1970,mahan_many-particle_2000,vos_rpa_2025}. However, at the small $q$ and small $\omega$ relevant for acoustic plasmons, such prescriptions can violate basic constraints such as the correct static screening limit (see Methods) and may yield unphysical results. The origin of the problem is that disorder does not simply broaden single-particle propagators: it also generates essential vertex corrections required by conservation laws~\cite{mahan_many-particle_2000}.

Here we treat the disorder vertex correction in the polarizability in a controlled way. We start from the full polarizability in terms of fermionic Matsubara frequencies $\omega_n$:
\begin{align}
	P({\bf q}, i \omega_n) = &
		\frac{1}{\beta} \sum_{m} \sum_{\bf k} \Lambda({\bf q}, i \omega_n; {\bf k}, i \omega_m) 
	\nonumber \\ &
		 \times \mathcal{G} ({\bf k}, i \omega_m)
			\mathcal{G} ({\bf k}+ {\bf q}, i \omega_m + i \omega_n),
	\label{Eq:Polarization_vertex_maintext}
\end{align}
where $P$ is the polarizability, $\mathbf{q}$ is the 3D momentum, $\mathcal{G}$ is the fully dressed electron Green's function, and $\Lambda$ is the vertex function~\cite{mahan_many-particle_2000}. Considering the small $q,\omega_n$ limit relevant for the acoustic plasmon and performing analytical continuation to real frequencies $i\omega_n\to \omega + i0^+$, we obtain the dielectric function for a two-band system:
\begin{equation}
    \epsilon({\bf q},\omega)= 1-\sum_{\nu=1}^{2}\frac{\omega_{\textrm{pl},\nu}^2}{\omega(\omega+i\Gamma)-\frac{1}{3}v_{\textrm{F},\nu}^2 q^2},
\label{eq:two_band_dielectric_function}
\end{equation}
where $\omega_{\textrm{pl},\nu}$ is the plasma frequency, and $v_{\textrm{F},\nu}$ the Fermi velocity, of band $\nu$. A detailed derivation of this dielectric function can be found in the Methods. 

Pines' demon requires the existence of one light and one heavy band~\cite{pines_demon_original}. To obtain the analytical form of the dispersion, we approximate the contribution of the light band by the Thomas-Fermi screening dielectric function, corresponding to $\omega \rightarrow 0$~\cite{mahan_many-particle_2000}. For the heavy band we instead take the long-wavelength limit $q\to0$, resulting in the two-band dielectric function
\begin{equation}
\label{eq:approximation_dielectric_function_two_bands}
    \epsilon_{2-\mathrm{band}}(q,\omega) = 1+\frac{q_{\textrm{TF,L}}^2}{q^2} - \frac{\omega_{\textrm{pl,H}}^2}{\omega(\omega+i\Gamma)},
\end{equation}
where $q_{\textrm{TF,L}}$ is the Thomas-Fermi screening wave vector of the light band and $\omega_{\textrm{pl,H}}^2$ is the plasma frequency of the heavy band. We find the acoustic plasmon dispersion by solving $\epsilon_{2-\mathrm{band}}(q,\omega) = 0$, giving
\begin{equation}
\label{eq:dispersion_approximation_two_bands}
    \omega_{2-\mathrm{band}}(q,\omega) = -i\frac{\Gamma}{2} \pm \sqrt{q^2 \frac{\omega_{\textrm{pl,H}}^2}{q_{\textrm{TF,L}}^2} -\frac{\Gamma^2}{4}}.
\end{equation}
The hydrodynamic result of Eq.~\eqref{eqn:hydro_dispersion} is recovered for $v_s= \omega_{\textrm{pl,H}}/q_{\textrm{TF,L}}$.

Working directly with the dielectric function in Eq.~\eqref{eq:two_band_dielectric_function}, we can find the zeroes of $\epsilon({\bf q},\omega)=0$ numerically without the approximations leading to Eq.~\eqref{eq:dispersion_approximation_two_bands}. We obtained the parameters in Eq.~\eqref{eq:two_band_dielectric_function} by fitting a two-band parabolic model to the tight-binding model for Sr$_2$RuO$_4$ and scaling the Fermi velocities with temperature, following the procedure used for the hydrodynamic estimates and detailed in the Methods. Owing to the quadratic band approximation, the resulting dispersion (Fig.~\ref{fig:main_results_dispersion}) matches the result obtained with hydrodynamics.

The loss function, defined as~\cite{mahan_many-particle_2000,husain_pines_2023} $S(q,\omega)=-\mathrm{Im}(\epsilon(q,\omega)^{-1})$, encodes information about the real and imaginary parts of the dielectric function for real values of $\omega$. It indicates where the acoustic mode is expected to appear in the $(q,\omega)$ plane. The damping is shown as a heatmap in Fig.~\ref{fig:main_results_dispersion}: brighter regions indicate a stronger predicted signal of the demon as measured for example in m-EELS. We observe that the brighter regions coincide with the experimental data, providing a further confirmation that the mode is well-defined in that frequency and momentum regime in Sr$_2$RuO$_4$ and that the measurements in Ref.~\cite{husain_pines_2023} correspond to a momentum-relaxed demon.
    
\section{Conclusions}

Hydrodynamics and its extension to slowly-relaxing momentum has become a vital tool in the study of electronic systems in which entanglement and strong correlations cause the breakdown of the quasiparticle picture and the failure of standard microscopic theories. Notable examples include the `strange metals' from which many cuprate high-Tc superconductors form, `bad metals' which include Sr$_2$RuO$_4$, and graphene~\cite{hartnoll_holographic_2018,lucas_hydrodynamics_2018,phillips_stranger_2022}. Degree-of-freedom counting often suggests a connection between macroscopic hydrodynamic modes and microscopic quasiparticle dispersions, but the direct demonstration of an adiabatic connection between the two has so far proven elusive even in simple models. This makes the exact analytic agreement we report here -- between hydrodynamics on the one hand, and quasiparticle dispersions of disordered multi-band metals on the other -- all the more remarkable. It seems reasonable to hope that this connection may provide clues in the search for microscopic descriptions of systems in which hydrodynamics provides the best current understanding.

Our results also reveal a general feature of quasiparticle dispersions. Any real material has momentum relaxation -- be it from defects, disorder, or finite crystal size -- and so \emph{any} putative gapless bosonic quasiparticle will feature a momentum relaxation scale. In this sense, the detailed dispersion of Pines' demon we report in Fig.~\ref{fig:main_results_dispersion} is generic. There are two main reasons such dispersions had not previously been measured. First, emergent bosons' masslessness is often symmetry--protected, such as when they arise from Goldstone's theorem (as in the case of acoustic phonons in crystals) or gauge constraints. Second, a mass gap will also typically be present; if the energy of the gap is larger than the momentum relaxation energy scale, the familiar massive-particle dispersion will result. Importantly, a mass gap will always open in the presence of Coulomb interactions: the standard optical plasmon is an example. The special feature of Pines' demon that allowed its novel dispersion to be revealed is that it is charge neutral. This made its experimental observation all the more remarkable~\cite{husain_pines_2023}. Nevertheless, there exist other candidates to show momentum-relaxed dispersions, including Leggett modes in multiband superconductors~\cite{10.1143/PTP.36.901} and in structural phase transitions~\cite{PhysRevX.12.011024}.

Crucially, acoustic plasmons could play a role in the superconducting pairing of exotic and high-Tc superconductors, such as metal hydrides~\cite{Akashi_density_2014,pashitskii_possibility_2022}, by lowering the Coulomb repulsion between electrons~\cite{ruvalds1979superconductivity,ruvalds_disorder_1979,tutto_tunneling_1979,kliewer_lindhard_1969,ihm_demons_1981}. In this regard, it is relevant to ask if the demon plays a role in the exotic but not yet fully understood multiband superconductivity of Sr$_2$RuO$_4$~\cite{Tyler_high_1998,Mackenzie_superconductivity_2003,Wang_quasiparticle_2004,stricker_optical_2014,Wagner21,Roising19}, and the high-T$_c$ superconductivity of multiband metals.

\section*{Acknowledgments}

The authors thank Peter Abbamonte for detailed feedback on the manuscript, Dirk van der Marel for providing the optical conductivity data used in this work, and Christophe Berthod, Luke C.~Rhodes and Peter Wahl for helpful discussions and related collaborations. M.-Á.~S.-M.~and F.~F.~acknowledge support from the Engineering and Physical
Sciences Research Council, Grant No.~EP/X012239/1.
L.~R.~acknowledges the Swiss National Science Foundation (SNSF) via Starting Grant TMSGI2 211296. 

\section{Methods}

\subsection{Hydrodynamics}

In this section, we show how two independent bands of charged particles give rise to a neutral acoustic mode when treated hydrodynamically. We model the bands by two densities coupling only through  the external electric fields
\begin{align}
    \label{eqn:hydro_conservation_equations2}
    \partial_t \pi_a^i + n_a\partial^i\mu_a&=-\Gamma_a n_a v_a^i-e \frac{n_a}{m_a}E^i,\\ \nonumber
    \partial_tn_a + \partial_i (n_a v_a^i)&=0,\nonumber
\end{align}
where $a=1,2$ (no sum over $a$). For simplicity, we assume there is no interband momentum transfer. We also set to zero thermodynamic cross-derivatives such as $\partial n_1/\partial\mu_2$, and similarly momentum drag of one species by the other $\partial\pi^i_1/\partial v_2^i=0$, so that $\pi_a^i=n_a v_a^i$ and $j_a^i=-en_a v_a^i$ for the charge current. 

We perturb the hydrodynamic fields $\Psi=(n_a,\mu_a,v_a^i)$ around a local homogeneous thermal equilibrium, $\Psi = \bar \Psi+\delta \Psi(\omega,q) e^{-i\omega t+i q x}$, with $\bar \Psi=(\bar n_a,\bar\mu_a,\vec 0)$ the equilibrium values. To close the system of equations, we use $\delta n_a=(\partial n_a/\partial \mu_a)_{T,\mu_{b\neq a}}\delta \mu_a$. This allows us to solve for all $\delta\Psi$ in terms of the electric field $E^i$. We can compute the local conductivity from 
\begin{equation}
\label{eqn:hydroconductivity}
    \sigma_{ij}(\omega,k)=\frac{\delta j_1^i}{\delta E^j}+\frac{\delta j_2^i}{\delta E^j}=\sigma(\omega,k)\delta_{ij}\,,
\end{equation}
since both charge currents couple to the same external electric field. The last equality follows from the isotropy of the thermal equilibrium state and the absence of parity violation. Since both fluids are treated on equal footing, their individual conductivities are
\begin{equation}
\label{eqn:hydroconductivity2}
    \sigma_a(\omega,k)=\frac{n_a e^2}{m_a}\frac{i\omega}{\omega^2+i\omega\Gamma_a-v_{s,a}^2q^2}
\end{equation}
where $v_{s,a}^2=n_a/(\partial n_a/\partial\mu_a)_{T,\mu_{b\neq a}}$. Each band carries collective excitations in the form of relaxed sound poles, as described in equation \eqref{eqn:hydro_dispersion}. 

In the presence of long-range Coulomb interactions, the collective excitations of the charge susceptibility are instead given by the zeroes of the dielectric function
\begin{align}
	\epsilon(\omega,q) = 1 -  \frac{\sigma(\omega,q)}{i \omega \epsilon_0}.
\end{align}
Using \eqref{eqn:hydroconductivity} and \eqref{eqn:hydroconductivity2}, this gives a quartic equation whose solutions can be worked out explicitly. It is more instructive to expand them for small wavenumber $q$ and small $\Gamma$ (in this order of limits):
\begin{align}
    \omega_P^\pm=&\pm e\sqrt{\frac{m_2 n_1+m_1 n_2}{m_1m_2\epsilon_0}}-\frac{i}{2}\Gamma+\mathcal O(q^2,\Gamma_a^2)\\
    \omega_{ac}^+=&-i\Gamma+i\frac{m_1n_2 v_{s1}^2+m_2n_1 v_{s2}^2}{m_2 n_1+m_1n_2}\frac{q^2}{\Gamma}\\
    \omega_{ac}^-=&-\frac{m_1n_2 v_{s1}^2+m_2n_1 v_{s2}^2}{m_2 n_1+m_1n_2}\frac{q^2}{\Gamma}.
\end{align}
Here $\left(\omega_P^{\pm}\right)^2=\omega_{P,1}^2+\omega_{P,2}^2$ are the optical plasmons broadened by disorder, while $\omega_{ac}^{\pm}$ are the relaxed acoustic plasmons. $\Gamma$ is the effective relaxation rate, given by the weighted sum of each band's momentum relaxation rate:
\begin{align}
\Gamma=&\frac{\Gamma_2\omega_{P,1}^2+\Gamma_1\omega_{P,2}^2}{\omega_{P,1}^2+\omega_{P,2}^2}
\end{align}
Since the small-$q$ limit is taken first, distances are long compared to the mean free path $\sim1/\Gamma$ and momentum is relaxed at $\omega=-i\Gamma$ while density diffuses quadratically. If we take the limits in the opposite order we instead find 
\begin{equation}
    \omega_{ac}^\pm=-\frac{i}{2}\Gamma\pm\sqrt{\frac{m_1n_2 v_{s1}^2+m_2n_1 v_{s2}^2}{m_2 n_1+m_1n_2}}q+\mathcal O(\Gamma^2,q^2)\,.
\end{equation}
This reveals that they are in fact acoustic plasmons, with a purely linear dispersion in the limit of vanishing $\Gamma$. 

\subsection{Diagrammatics}

From a microscopic perspective, the dielectric function $\epsilon({\bf q},\omega)$ can be calculated using linear response theory,
\begin{align}
	\epsilon({\bf q},\omega) = 1 - V({\bf q}) P({\bf q}, \omega),
	\label{Eq:DielectricPolarization}
\end{align}
where $P ({\bf q},i\omega_n)$ is the polarization function and $V({\bf q})$ the bare Coulomb repulsion. Within the random phase approximation (RPA), and in the absence of disorder, the polarization of a single band with dispersion $\xi_{\bf k}$ is approximated by the `bare' polarization
\begin{align}
	P_0 (\mathbf{q}, \omega) = \sum_{\bf k} \frac{n_F(\xi_{\bf k}) - n_F(\xi_{{\bf k}+{\bf q}}) }{\omega + \xi_{\bf k} - \xi_{{\bf k} + {\bf q}} + i 0^+},
	\label{Eq:Lindhard}
\end{align}
also known as the Lindhard function. The long-wavelength limit $q \rightarrow 0$ at finite frequency $\omega >0$ yields $P_0 (q \ll \omega) = \frac{n q^2}{m \omega^2}$ and this is the source of the {\em optical} plasmon, 
\begin{align}
	\epsilon_{\rm RPA} (q \ll \omega) = 1 - \frac{ne^2}{m \epsilon_0 \omega^2} = 1 - \frac{\omega_{\rm pl}^2}{\omega^2}.
\end{align}

Let us now try to include the effects of disorder. On the level of the electron Greens function, disorder yields (using the self-consistent Born approximation) an imaginary part of the electron self-energy, which is conventionally written as $\text{Im} \Sigma = - \Gamma/2$. For the polarization, this suggests we can replace $\omega \rightarrow \omega + i \Gamma$. The long-wavelength limit of the dielectric function in this `RPA+$\Gamma$' approximation is
\begin{align}
	\epsilon_{\rm RPA + \Gamma} (q \ll \omega; \Gamma) = 1 - \frac{\omega_{\rm pl}^2}{(\omega + i \Gamma)^2}.
	\label{Eq:IncorrectDielectric}
\end{align}
This is result is {\em incorrect}. This can be seen by relating the long-wavelength dielectric function to the optical conductivity $\sigma(\omega)$
\begin{align}
	\epsilon(q =0, \omega) = 1 -  \frac{\sigma(\omega)}{i \omega \epsilon_0 }
\end{align}
and provided that the optical conductivity has the Drude form
\begin{align}
	\sigma(\omega) = \frac{ne^2}{m} \frac{1}{\Gamma - i \omega}
\end{align}
the dielectric function in the limit $q \rightarrow 0$ should have the shape
\begin{equation}
	\epsilon(\omega) = 1 - \frac{\omega_{\rm pl}^2}{\omega (\omega + i \Gamma)}.
	\label{Eq:DielectricWithGammaCorrect}
\end{equation}
Physically, the problem of the approximated form Eq.~\eqref{Eq:IncorrectDielectric} is that in the limit $\omega \rightarrow 0$ the inverse $1/\epsilon$ should vanish for a metal~\cite{mermin_lindhard_1970}. Note that Eq.~\eqref{Eq:DielectricWithGammaCorrect} does satisfy this physical constraint.

Since the `RPA+$\Gamma$' approximation is incorrect, we need to go beyond RPA by including vertex corrections.
Formally, the full polarization in terms of Matsubara frequencies $\omega_n$ can be expressed as
\begin{align}
	P({\bf q}, i \omega_n) = &
		\frac{1}{\beta} \sum_{m} \sum_{\bf k} \Lambda({\bf q}, i \omega_n; {\bf k}, i \omega_m) 
	\nonumber \\ &
		 \times \mathcal{G} ({\bf k}, i \omega_m)
			\mathcal{G} ({\bf k}+ {\bf q}, i \omega_m + i \omega_n)
	\label{Eq:FormalPolarizationwithvertex}
\end{align}
where $\mathcal{G}$ is the fully dressed electron Greens function and $\Lambda$ is the vertex function. We use the common approximation to express the vertex function self-consistently using the Bethe-Salpeter equation. For momentum-independent impurity scattering, the vertex has no dependence on the electron momentum and the Bethe-Salpeter equation reads
\begin{align}
	\Lambda({\bf q}, i \omega_n; i \omega_m) =& 1 +  \Lambda({\bf q}, i \omega_n;  i \omega_m)  \frac{\Gamma}{2\pi N_0} \nonumber \\ &
	\times \sum_{\bf k'}  
	 \mathcal{G} ({\bf k'}, i \omega_m)
	\mathcal{G} ({\bf k'}+ {\bf q}, i \omega_m + i \omega_n)
\end{align}
where $N_0$ is the bare electron density of states at the Fermi level and $\Gamma$ is the momentum relaxation rate due to disorder. The vertex function is thus
\begin{align}
	 & \Lambda({\bf q}, i \omega_n; i \omega_m) = \nonumber \\ &
		\frac{1}{1 - \frac{\Gamma}{2\pi N_0} \sum_{\bf k'}  
		 \mathcal{G} ({\bf k'}, i \omega_m)
			\mathcal{G} ({\bf k'}+ {\bf q}, i \omega_m + i \omega_n)}.
	\label{Eq:VertexFunctionBetheSalpeterSolution}
\end{align}
Let us define the part with summation over ${\bf k'}$ as
\begin{align}
	\tilde{P} ({\bf q}, i\omega_n; i \omega_m)
	= \sum_{\bf k}
		 \mathcal{G} ({\bf k}, i \omega_m)
			\mathcal{G} ({\bf k}+ {\bf q}, i \omega_m + i \omega_n).
\end{align}
For a parabolic dispersion $\xi_{\bf k} = \frac{k^2}{2m} - E_F$, with $k=|\mathbf{k}|$, in $d=3$ dimensions, this becomes
\begin{align}
   & \tilde{P}
    = \int \frac{\textrm{d}^3\mathbf{k}}{(2\pi)^3} 
        \frac{1}{i \omega_m - \xi_{\bf k} + i \frac{\Gamma}{2} {\rm sgn} (\omega_m)} 
        \nonumber \\ &
        \times \frac{1}{i \omega_m + i \omega_n - \xi_{\bf k} - \frac{kq}{m} \cos \theta - \frac{q^2}{2m} + i \frac{\Gamma}{2} {\rm sgn} (\omega_m+\omega_n).}
\end{align}
The integral over $|{\bf k}|$ can be replaced with an integral over the dispersion $\xi_{\bf k}$. Since the Greens functions are highly peaked near the Fermi level, we extend the integral over $\xi$ to all values of $\xi$, set the density of states $N(\xi)$ equal to that at the Fermi level $N(\xi) \rightarrow N_0$,  and replace any instance of $|{\bf k}| \equiv k$ with $k_F$. This approximation, valid for $\omega, \Gamma \ll E_F$, yields
\begin{align}
    &\tilde{P} 
    =
    N_0 \int \frac{\sin \theta \textrm{d}\theta \textrm{d}\phi}{4 \pi} \int d\xi 
        \frac{1}{i \omega_m - \xi + i \frac{\Gamma}{2} {\rm sgn} (\omega_m)}
    \nonumber \\ & \times
        \frac{1}{i \omega_m + i \omega_n - \xi - v_F q \cos \theta - \frac{q^2}{2m} + i \frac{\Gamma}{2} {\rm sgn} (\omega_m+\omega_n)}
\end{align}
where $v_F = k_F/m$. The integral over $\xi$ can now be performed using standard contour integration methods. For $\omega_n >0$, the integral only yields a nonzero result when $\omega_m + \omega_n < 0$, because then the two poles are in the opposite half-planes. We thus get
\begin{align}
    \tilde{P} =&
    \Theta(-\omega_n - \omega_m)
    N_0 \int \frac{\sin \theta \textrm{d}\theta \textrm{d}\phi}{4 \pi} \nonumber \\ & \times
    \frac{2\pi i}{i \omega_n- v_F q \cos \theta - \frac{q^2}{2m} + i \Gamma}.
\end{align}
Finally, we expand the integrand quadratically in momentum $q$, ignore the imaginary component of the vertex, and insert the result for $\tilde{P}$ into Eq.~\eqref{Eq:VertexFunctionBetheSalpeterSolution} to obtain the vertex function
\begin{align}
	    \Lambda =
        \left\{
        \begin{matrix}
            \left(1 - \frac{\Gamma}{\Gamma + \omega_n}
    + \frac{v_F^2 \Gamma}{3 (\Gamma + \omega_n)^3} q^2\right)^{-1} & {\rm if } - \omega_n < \omega_m < 0, \\
            1 & {\rm otherwise}.
        \end{matrix}
        \right.
    \label{Eq:VertexCorrection80}
\end{align}
We have a vertex correction $\Lambda({\bf q},i\omega_n; i \omega_m)$ that does not depend on $\omega_m$, with the exception of the Heaviside functions. We thus approximate our final polarization function by the vertex correction times the Lindhard function corrected by disorder,
\begin{align}
    P_{\rm RPA + \Gamma + \Lambda}({\bf q}, i \omega_n) \approx
    \Lambda({\bf q}, i\omega_n) P_{0} ({\bf q}, i \omega_n + \Gamma).
\end{align}
In the optical range $q \ll \omega_n$, the `shifted' Lindhard function is
\begin{align}
    P_{0} ({\bf q}, i \omega_n + \Gamma)
        = - \frac{n q^2}{m(\omega_n + \Gamma)^2}.
\end{align}
Note that the condition $q \ll \omega_n$ implies that momentum is small compared to temperature -- precisely the regime that is described by hydrodynamics. 
Including the vertex correction of Eq.~\eqref{Eq:VertexCorrection80} yields
\begin{align}
    P_{\rm RPA + \Gamma + \Lambda}({\bf q}, i \omega_n)
    = - \frac{nq^2/m}{\omega_n (\omega_n + \Gamma) + \frac{v_F^2 \Gamma}{3 (\Gamma + \omega_n)} q^2}.
\end{align}
Expanding for small $q, \omega_n$, we find $\frac{q^2}{\Gamma + \omega_n} \approx q^2 / \Gamma$ and thus
\begin{align}
    P_{\rm RPA + \Gamma + \Lambda}({\bf q}, i \omega_n)
    = - \frac{nq^2/m}{\omega_n (\omega_n + \Gamma) + \frac{1}{3} v_F^2 q^2}.
\end{align}
Analytical continuation to real frequencies, $i\omega_n\rightarrow\omega+i0^+$, yields
\begin{align}
    P_{\rm RPA + \Gamma + \Lambda}({\bf q}, \omega)
    = \frac{nq^2/m}{\omega (\omega  + i \Gamma) - \frac{1}{3} v_F^2 q^2}.
\end{align}
It is instructive to compare this polarization to the hydrodynamic result of Eq.~\eqref{eqn:hydroconductivity2}: they are identical if we associate $v_{s,a}^2 = \frac{1}{3} v_F^2$, which is precisely the known result for the velocity of hydrodynamic sound of an isotropic Fermi liquid in $d=3$, \cite{landau2013statistical}, in the absence of Landau marginal interactions. 

The corresponding dielectric function reads
\begin{align}
	\epsilon({\bf q}, \omega) = 1 - \frac{\omega_{\rm pl}^2}{\omega (\omega + i \Gamma) -\frac{1}{3} v_F^2 q^2 }
	\label{Eq:RPAGammaDielectricFinal}.
\end{align}
The static limit of the dielectric function, which is unaffected by the vertex corrections and disorder, is recovered by the above expression for $\epsilon(q,\omega)$.

The case of a multi-band system can be described by combining Eq.~\eqref{Eq:RPAGammaDielectricFinal} for each band. For each band, we just need to know the momentum relaxation rate $\Gamma$, the Fermi velocity $v_F$, and the optical plasmon frequency $\omega_{\rm pl}$. 
For example, an acoustic plasmon is captured by the standard Thomas-Fermi screening result for the light band, $\omega\rightarrow 0$, and the long-wavelength limit $q\rightarrow 0$ of Eq.~\eqref{Eq:RPAGammaDielectricFinal} for the heavy band. Combining, this yields for a two-band system
\begin{equation}
	\epsilon_{\text 2-band} (q,\omega) = 
	1 + \underbrace{\frac{q_{{\rm TF},L}^2}{q^2}}_{\rm light \, band}
	- \underbrace{\frac{\omega_{{\rm pl}, H}^2}{\omega(\omega+ i\Gamma)} }_{\rm heavy \, band}.
\end{equation}
Solving for $\epsilon(q,\omega) = 0$ yields the momentum-relaxed acoustic plasmon mode. 

\subsection{Fermi velocities and scaling with temperature}

We will work with a model consisting of two parabolic bands of the form 
\begin{equation}
    H_{\mathrm{2-band}}(\mathbf{k})=\mathrm{diag}\left(\frac{\mathbf{k}^2}{2m_L}-\mu_L,\frac{\mathbf{k}^2}{2m_H}-\mu_H\right),
    \label{eq:two_band_hamiltonian}
\end{equation}
where $m_L$ ($m_H$) and $\mu_L$ ($\mu_H$) are the mass and chemical potential of the electrons in the light (heavy) band, respectively. 

To obtain the plasma frequencies $\omega_{pl,\nu}$ and Fermi velocities $v_{F,\nu}$ $(\nu=L,H)$ in the dielectric function in Eqs.~\eqref{eq:two_band_dielectric_function} and \eqref{eq:approximation_dielectric_function_two_bands}, and the corresponding acoustic plasmon dispersion in Eq.~\eqref{eq:dispersion_approximation_two_bands}, we fit the two-band model in Eq.~\eqref{eq:two_band_hamiltonian} to the tight-binding model capturing the relevant bands of Sr$_2$RuO$_4$ close the Fermi level presented in Ref.~\cite{zabolotnyy_renormalized_2013}. The acoustic plasmon in Sr$_2$RuO$_4$ is composed mainly of electrons in the $\beta$ and $\gamma$ bands~\cite{husain_pines_2023}; therefore, we will estimate the parameters of the parabolic bands in Eq.~\eqref{eq:two_band_hamiltonian} to capture the behaviour of the $\beta$ and $\gamma$ bands in Sr$_2$RuO$_4$. 

\begin{figure}[h!]
    \centering
    \includegraphics[width=\linewidth]{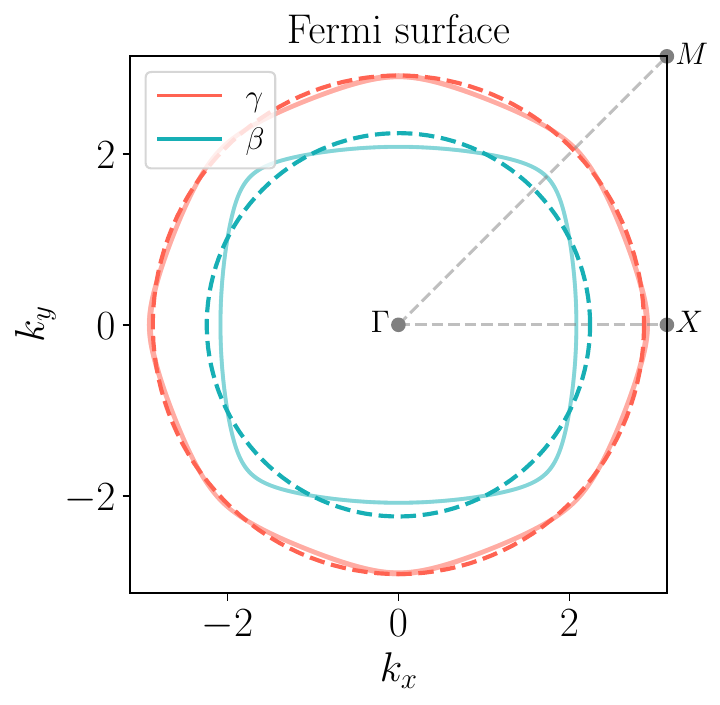}
    \caption{Fit of the parabolic bands to the relevant bands of the tight-binding model of Sr$_2$RuO$_4$ in Ref.~\cite{zabolotnyy_renormalized_2013}. The Fermi surface of the tight-binding model is indicated in solid colors for the $\gamma$ (red) and $\beta$ (blue) bands. The Fermi surface of the parabolic bands is indicated by dashed lines. The high-symmetry directions $\Gamma X$ and $\Gamma M$ of the tight-binding model are indicated, where $\Gamma=(0,0)$, $M=(\pi,\pi)$, and $X=(\pi,0)$.}
    \label{fig:fermi_surface_fitting}
\end{figure}

The values for the masses of electrons in the $\beta$ and $\gamma$ bands are extracted from Refs.~\cite{bergemann_normal_2001,Bergemann_quasi_2003,zabolotnyy_renormalized_2013}, leading to $m_H=16~m_e$ and $m_L= 7~m_e$. To estimate the values of the chemical potentials $\mu_L, \mu_H$ we compute the average value of the Fermi waveve numbers $k_{F,\beta}$ and $k_{F,\gamma}$ in the tight-binding model in Ref.~\cite{zabolotnyy_renormalized_2013}. Since the masses are fixed in the model in Eq.~\eqref{eq:two_band_hamiltonian}, the chemical potentials are fixed to obtain a circular Fermi surface with the Fermi wave numbers estimated for the tight-binding model, leading to the values $\mu_H=0.26$ eV and $\mu_L=0.36$ eV and the circular Fermi surfaces presented in Fig.~\ref{fig:fermi_surface_fitting}.

The plasma frequencies $\omega_{pl,\nu}$, Fermi velocities $v_{F,\nu}$, and the Thomas-Fermi wavevector $q_{\mathrm{TF,\nu}}$ $(\nu=L,H)$ are given by the standard relations for parabolic dispersions~\cite{bruus_many-body_2004}:
\begin{eqnarray}
    \omega_{pl}&=&\sqrt{\frac{4k_F^3}{3\pi m}},\label{eqn:plasma_frequency}\\
    v_F&=& \frac{k_F}{m},\\
    q_{\mathrm{TF}}&=&\sqrt{4\pi N(E_F)},\\
    N(E_F)&=&\frac{1}{2\pi^2}\sqrt{(2m)^{3}\mu}, 
\end{eqnarray}
where $N(E_F)$ is the density of states at the Fermi level of the corresponding parabolic band.

\begin{figure}[h!]
    \centering
    \includegraphics[width=\linewidth]{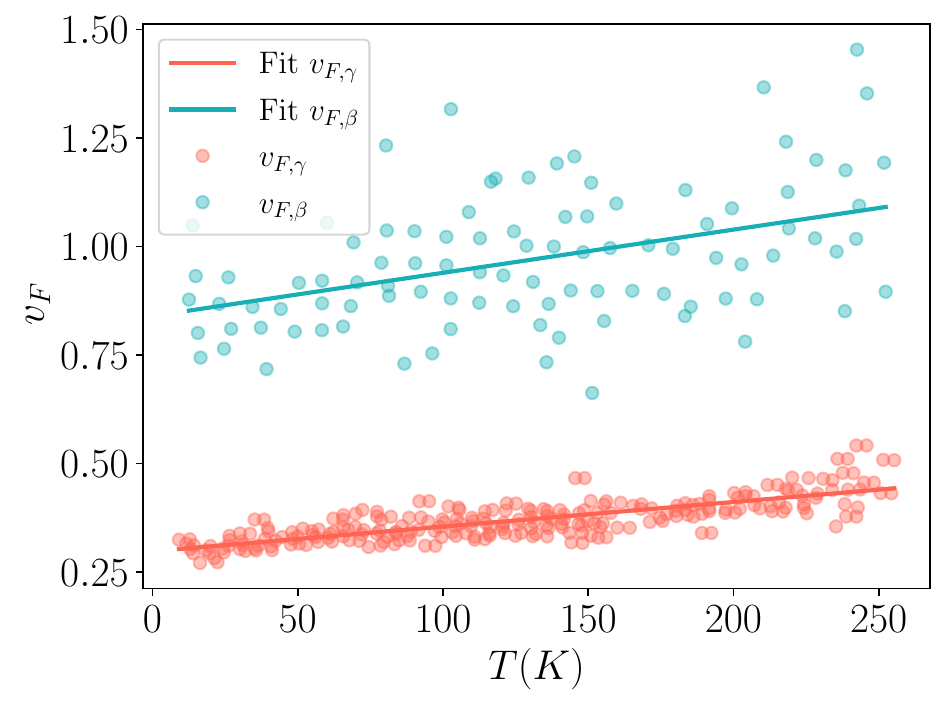}
    \caption{Fermi velocities of the $\beta$ (red) and $\gamma$ (blue) bands of Sr$_2$RuO$_4$ as a function of temperature. The circles correspond to the data reported in Ref.~\cite{hunter_fate_2023}, and the solid lines to the corresponding least squares linear fit.}
    \label{fig:linear_fit_temperature_scaling}
\end{figure}

To account for the temperature scaling of the Fermi velocities, we perform a linear fit of $v_{F,\beta}$ and $v_{F,\gamma}$ measured at different temperatures in Ref.~\cite{hunter_fate_2023}, presented in Fig.~\ref{fig:linear_fit_temperature_scaling}. We obtain a linear dependence $v_{F,\beta}(T)=v_{0,\beta}+s_{\beta}T$, with a linear scaling $s_{\beta}=9.955\times10^{-4}\,\mathrm{eV}\,\text{\AA}\,\mathrm{K}^{-1}$ and an intercept at $T=0$~K of $v_{0,\beta}=0.8397\,\mathrm{eV}\,\text{\AA}$, with $R^2=0.1924$ for the $\beta$ band. For the $\gamma$ band we have $v_{F,\gamma}(T)=v_{0,\gamma}+s_{\gamma}T$, with a linear scaling $s_{\gamma}=5.676\times10^{-4}\,\mathrm{eV}\,\text{\AA}\,\mathrm{K}^{-1}$ and an intercept at $T=0$~K of $v_{0,\gamma}=0.2978\,\mathrm{eV}\,\text{\AA}$, with $R^2=0.6237$. This leads to a ratio between the Fermi velocities at $T=30$~K and $T=300$~K of ${v_{F,\beta}(30{\rm K})}/{v_{F,\beta}(300{\rm K})}=0.6726$, and ${v_{F,\gamma}(30{\rm K})}/{v_{F,\gamma}(300{\rm K})}=0.7639$.

The Fermi velocities measured at high temperatures are closer to the bare Fermi velocities~\cite{hunter_fate_2023} estimated above for the Hamiltonian in Eq.~\eqref{eq:two_band_hamiltonian}. Therefore, we take $v_{F,\gamma}(300)$ and $v_{F,\beta}(300)$ as those estimated from the fit to the tight-binding model, and obtain $v_{F,\gamma}(30)$ and $v_{F,\beta}$(30) from the temperature scaling.

The hydrodynamic dispersion requires an estimation of the speed of sound in the system. To do so, we take the dielectric function in Eq.~\eqref{eq:two_band_dielectric_function} in the limit $\Gamma\to 0$, which results in a linear mode with velocity $v_S(300\mathrm{K})=0.148\,\mathrm{eV}\,\text{\AA}$ at $T=300$~K, and $v_S(30\mathrm{K})=0.103\,\mathrm{eV}\,\text{\AA}$ at $T=30$~K.

\subsection{Estimation of $\Gamma$}

Momentum relaxation in metallic systems contributes to the optical conductivity at low frequencies through the Drude conductivity, which for a parabolic band structure reads~\cite{bruus_many-body_2004}:
\begin{equation}
    \label{eqn:drude_complex_conductivity}
    \sigma(\omega) = \epsilon_0\frac{\omega_{pl}^2}{\Gamma-i\omega},
\end{equation}
where $\epsilon_0$ is the vacuum permittivity, $\omega_{pl}$ is the plasma frequency, $\Gamma$ is the momentum relaxation rate $\Gamma=1/\tau$, $\tau_{mr}$ is the momentum relaxation time, and $\omega$ is the frequency of the applied field.

\begin{figure}
    \centering
    \includegraphics[width=\linewidth]{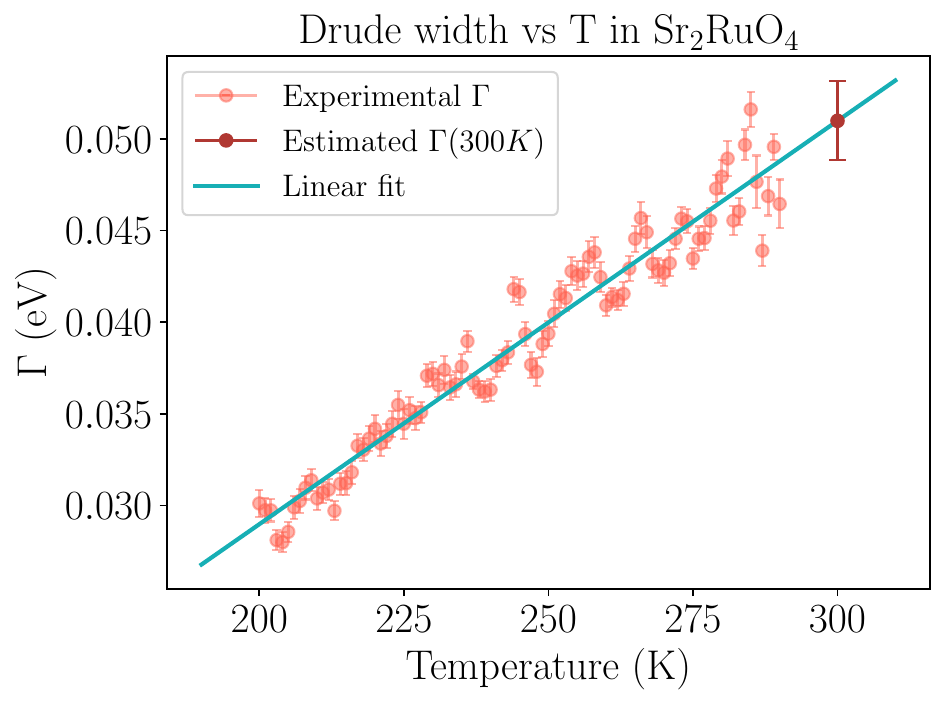}
    \caption{Estimation of $\Gamma$ from the Drude peak of optical conductivity measurements in Ref.~\cite{stricker_optical_2014} (light red circles) with $\omega_{max}=0.15$~eV using Eq.~\eqref{eq:Drude_fit} with their respective standard errors. The linear fit (blue) is used to estimate the value of $\Gamma$ at $T=300$~K (dark red circle) $\Gamma(300{\rm~K})=51\pm 2$~meV.}
    \label{fig:gamma_estimation}
\end{figure}

The optical conductivity of Sr$_2$RuO$_4$ has been measured over a wide range of temperatures~\cite{stricker_optical_2014}, where the low-energy response is well captured with two Drude oscillators. At small frequencies and temperatures above $40$~K, the conductivity can be fitted with a single Drude oscillator~\cite{wang_separated_2021}. Therefore, to obtain the momentum relaxation scale at $T=300$~K used in momentum-relaxed hydrodynamics in Sec.~\ref{sec:hydrodynamics} and in the microscopic theory in Sec.~\ref{sec:microscopic}, we fit the real part of the measured optical conductivity reported in Ref.~\cite{stricker_optical_2014} with a Drude function of the form:

\begin{equation}    \sigma^D(\omega)=\frac{\epsilon_0\omega_{pl}^2\Gamma}{\omega^2+\Gamma^2}+\sigma^{0},
\label{eq:Drude_fit}
\end{equation}
where $\sigma_0$ is the background contribution from the tails of other Lorentz oscillators at finite frequencies~\cite{stricker_optical_2014}. The resulting value of $\Gamma$ obtained from the experimental optical data is shown in Fig.~\ref{fig:gamma_estimation}.

\begin{figure}
    \centering
    \includegraphics[width=\linewidth]{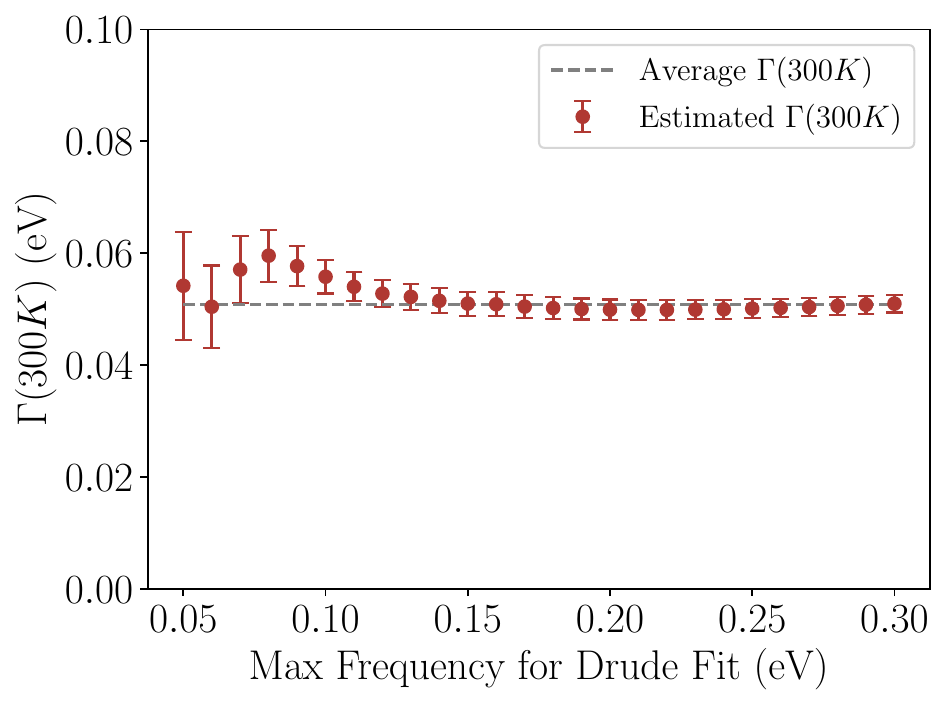}
    \caption{Estimated $\Gamma(300{\rm K})$ for different values of the maximum frequency $\omega_{max}$ used to fit the Drude peak. The weighted average value for $0.05~{\rm eV}<\omega_{max}<0.3~{\rm eV}$ is $\Gamma(300{\rm~K})=51$~meV.}
    \label{fig:gamma_300K_maxfreqs}
\end{figure}
We fit the optical data for temperatures between $T=200$~K and $T=290$~K, where the single-Drude fit is robust~\cite{wang_separated_2021} and the value of $\Gamma$ can be estimated reliably. The optical data used for the fit were originally reported in Ref.~\cite{stricker_optical_2014}, and provided by Dirk van der Marel for the present work. To estimate the value of $\Gamma$ at $T=300$~K we perform a linear fit of the values of $\Gamma$ between $T=200$~K and $T=290$~K, where the resistivity is $T$-linear~\cite{Berger_linear_1998}, resulting in a scaling with temperature of the form $\Gamma(T)=s_{\Gamma} T + \Gamma_0$, with $s_{\Gamma}=2.2\times10^{-4}\pm5.5\times10^{-6}$~eV K$^{-1}$ and $\Gamma_0=-0.015\pm0.0014$~eV, with $R^2=0.946$. As a result, the extrapolation to $T=300$~K, where no optical data are available, yields $\Gamma(300~\mathrm{K})=0.05098\pm0.00215$~eV. This result is consistent for different values of the maximum frequency used for the Drude fit (see Fig.~\ref{fig:gamma_300K_maxfreqs}). For low values of $\omega_{max}<0.08$~eV the estimation has a larger uncertainty, and leads to consistent results for $\omega_{max}>0.12$~eV, with an overall weighted average of $\Gamma(300{\rm K})=51$~meV, equal to the estimation presented in Fig.~\ref{fig:gamma_estimation} and used in the main text for the acoustic plasmon dispersion at $T=300$~K. This value of the momentum relaxation rate $\Gamma=1/\tau_{mr}$ is considerably smaller than the measurements for the quasiparticle lifetimes $\Gamma_{qp}=1/\tau_{qp}$ at $T=250$~K reported in Ref.~\cite{hunter_fate_2023}, and therefore much smaller than the extrapolations to $T=300$~K, consistent with the hydrodynamic regime where $\tau_{mr}\gg \tau_{qp}$.

\begin{figure}
    \centering
    \includegraphics[width=\linewidth]{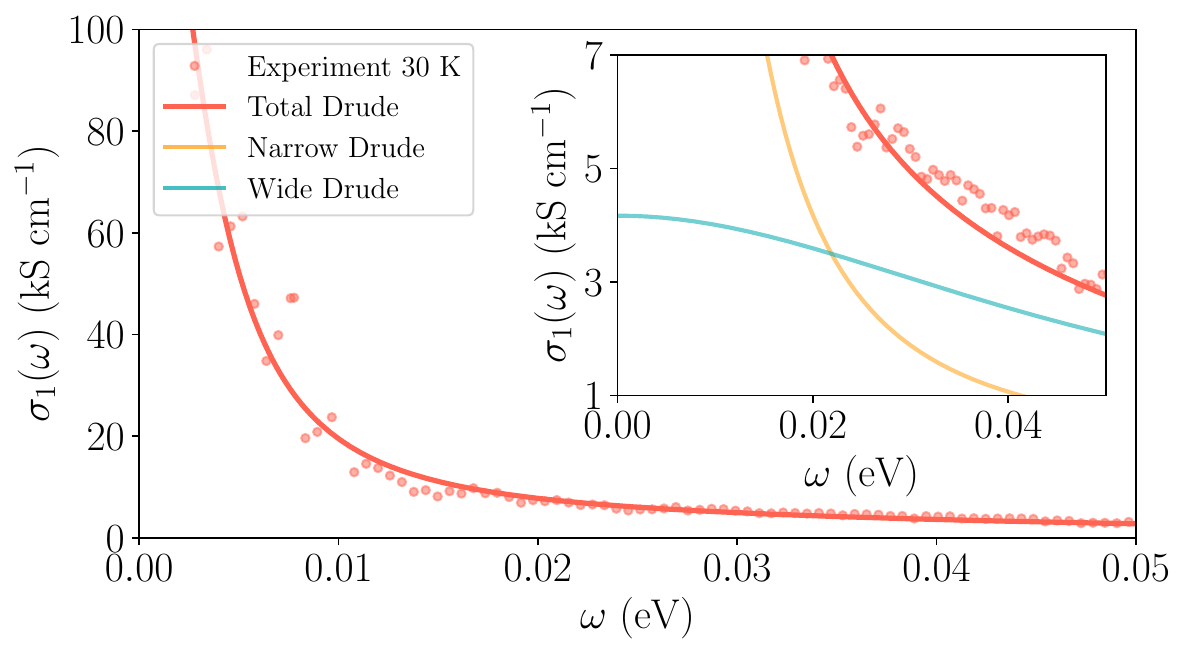}
    \caption{Drude fit to the optical conductivity measured at $T=30$ K (red circles) using the two-component Drude model in Eq.~\eqref{eqn:drude_two_components} (red line). The fit shows a narrow Drude (orange line, inset) with $\Gamma_1=3.207$~meV, and a wide Drude (blue line, inset) with $\Gamma_2=50$~meV.}
    \label{fig:drude_30K}
\end{figure}

At $T=30$~K, a single Drude fit cannot capture the behaviour of the optical conductivity for low frequencies~\cite{wang_separated_2021}. Therefore, we use a two-Drude model to fit the real part of the experimental optical conductivity~\cite{stricker_optical_2014}:
\begin{equation}
\sigma_2^D(\omega)=\epsilon_0\sum_{i=1,2}\frac{\omega_{pl,i}^2\Gamma_i}{\omega^2+\Gamma_i^2},
\label{eqn:drude_two_components}
\end{equation}
where $i=1,2$ corresponds to each of the peaks. At low frequencies, two Drude peaks are enough to account for all relevant scales in the optical conductivity~\cite{stricker_optical_2014}, and therefore we do not include a $\sigma_0$ term in the fit, unlike in the single-Drude case in Eq.~\eqref{eq:Drude_fit}. We perform a fit of Eq.~\eqref{eqn:drude_two_components} to the optical data with the only constraint being that the fitting parameters should be positive, and the value of the larger $\Gamma$ be less than or equal to the $300\,$K value. This leads to a good agreement with the experimental data (see Fig.~\ref{fig:drude_30K}), resulting in two Drude peaks with values of $\Gamma_1=3.26519\pm2\times10^{-5}$~meV and $\Gamma_2=50.0\pm4.8$~meV. We refer to the Drude peak with $\Gamma_1$ ($\Gamma_2$) as the narrow (wide) Drude. We note that Ref.~\cite{wang_separated_2021} highlights the need to anchor optical fits to a DC conductivity point. We did not do this with the data of Ref.~\cite{stricker_optical_2014}, due to the geometrical uncertainties they report (in Fig 1) in matching the AC data to the DC resistivity. 

The spectral weight $D_i$ of a single Drude is proportional to the plasma frequency squared $\omega_{pl,i}^2$~\cite{mahan_many-particle_2000}, given by Eq.~\eqref{eqn:plasma_frequency}. Therefore, the ratio of spectral weights of two Drude peaks is given by $D_{1}/D_{2}=\omega_{pl,1}^2/\omega_{pl,2}^2$. The value of the plasma frequency for the parabolic bands in Eq.~\eqref{eq:two_band_hamiltonian} is set by $\mu$ and $m$, leading to a ratio of spectral weights of both bands $D_{1}/D_{2}=2.483$ with the values of $m$ and $\mu$ used in this work. Our fit in Fig.~\ref{fig:drude_30K} leads to a ratio of spectral weights of $D_{1}/D_{2}=2.52$, which is in excellent agreement with the parabolic band picture used in the microscopic theory. Since we have shown that both theories lead to the same dispersion, the same momentum relaxation rate $\Gamma$ is used to obtain the dispersion in the hydrodynamic and microscopic theories.

\begin{figure}
    \centering
    \includegraphics[width=\linewidth]{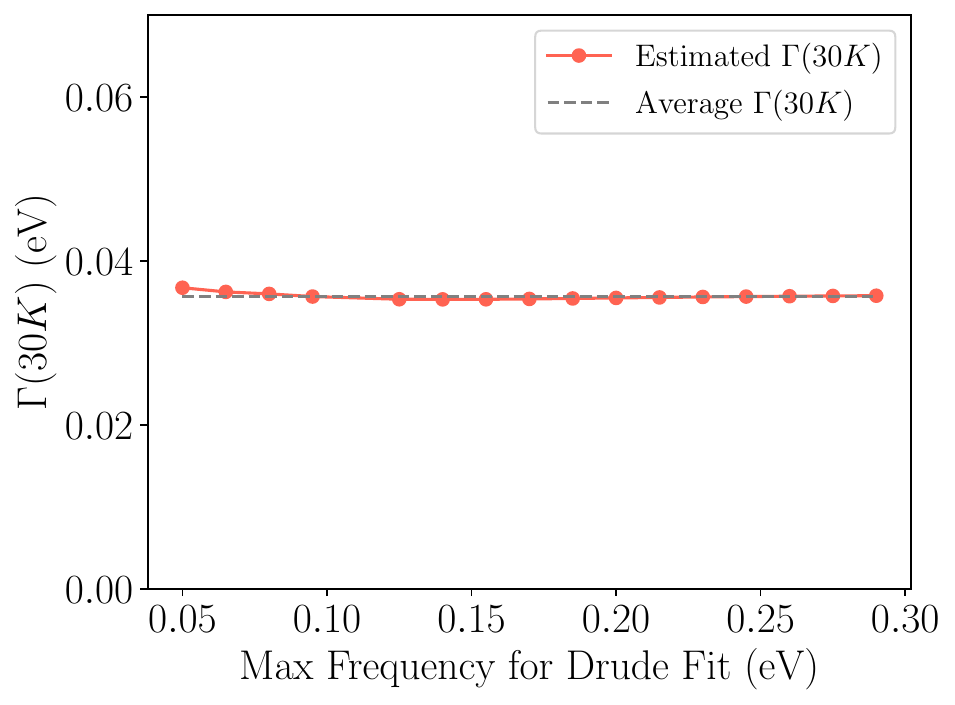}
    \caption{Estimated $\Gamma(30{\rm K})$ from the two Drude fit in Eq.~\eqref{eqn:drude_two_components} for different values of the maximum frequency $\omega_{max}$ used to fit the Drude peak. The estimated values deviate less than a $3\%$ from the average values of $\Gamma(30{\rm K})=35.67$~meV for $0.05~{\rm eV}<\omega_{max}<0.3~{\rm eV}$.}
    \label{fig:gamma_30K_maxfreqs}
\end{figure}

The single $\Gamma(30{\rm K})$ for the dispersion relation of the acoustic mode is given by the two-fluid model in Sec.~\ref{sec:hydrodynamics}. Such value can be estimated over a broad range of maximum frequencies for the fit of the Drude conductivity up to $\omega=300$~meV (see Fig.~\ref{fig:gamma_30K_maxfreqs}). The results are consistently close (less than a $3\%$ of difference) to an average value of $\Gamma(30~{\rm K})=35.67$~meV, which is the value used for the estimation of the dispersion of the acoustic plasmon at $T=30$~K in the main text.

\end{document}